# Spin-dependent transport and current modulation in a current-in-plane spin-valve field-effect transistor

Toshiki Kanaki,[1, a)] Tomohiro Koyama,[2] Daichi Chiba,[2] Shinobu Ohya,[1,3, b)] and Masaaki Tanaka[1,3, c)]

[1]*Department of Electrical Engineering and Information Systems, The University of Tokyo, 7-3-1 Hongo, Bunkyo-ku, Tokyo 113-8656, Japan*

[2]*Department of Applied Physics, The University of Tokyo, 7-3-1 Hongo, Bunkyo-ku, Tokyo 113-8656, Japan*

[3]*Center for Spintronics Research Network, Graduate School of Engineering, The University of Tokyo, 7-3-1 Hongo, Bunkyo-ku, Tokyo 113-8656, Japan*

We propose a current-in-plane spin-valve field-effect transistor (CIP-SV-FET), which is composed of a ferromagnet / nonferromagnet / ferromagnet trilayer structure and a gate electrode. This is a promising device alternative to spin metal-oxide-semiconductor field-effect transistors. Here, we fabricate a ferromagnetic-semiconductor GaMnAs-based CIP-SV-FET and demonstrate its basic operation of the resistance modulation both by the magnetization configuration and by the gate electric field. Furthermore, we present the electric-field-assisted magnetization reversal in this device.

[a)]Electronic mail: kanaki@cryst.t.u-tokyo.ac.jp

[b)]Electronic mail: ohya@cryst.t.u-tokyo.ac.jp

[c)]Electronic mail: masaaki@ee.t.u-tokyo.ac.jp

Spintronics, which utilizes not only the charge degrees of freedom but also the spin degrees of freedom in electronic devices, is a hopeful candidate for a next-generation technology. Various kinds of spintronics devices have been proposed thus far.[1–3] Above all, a spin metal-oxide-semiconductor field-effect transistor (spin MOSFET)[4,5], in which the source and drain electrodes are ferromagnetic materials, has attracted much attention because it is compatible with the sophisticated semiconductor technology and is expected to be used for non-volatile logic systems.[6] Although there have been some experimental demonstrations of both lateral and vertical spin MOSFETs,[7–10] there still remain problems to overcome for practical applications in the present status. In the lateral spin MOSFETs, the small magnetoresistance (MR) ratio (0.1%,[7] 0.005%,[8] and 0.03%[9]) is a crucial problem. This is caused by the difficulty of the efficient spin injection from ferromagnetic metals into semiconductors and of the coherent spin transport with a long distance in semiconductor channels. In the vertical spin MOSFETs, although the MR ratio is large, the current controllability is small in the present status and improvement is necessary.

In this Letter, we propose an alternative device, a current-in-plane spin-valve field-effect transistor (CIP-SV-FET) shown in Fig. 1(a). This device comprises a ferromagnet / nonmagnet / ferromagnet trilayer channel and a gate electrode, and can offer a functionality similar to spin MOSFETs. The resistance of the channel is controlled by the magnetization configuration through the current-in-plane spin-valve effect induced by the two ferromagnetic layers in the trilayer channel. The channel resistance is also controlled by the gate electric-field effect on the top ferromagnetic layer. Ideally, we can expect that the current is switched on/off by applying the gate electric field. Thus, we can use this device as a transistor. Moreover, if the thickness of the top ferromagnetic film is sufficiently thin, the magnetic properties can be modulated by the gate electric field.[11–13] Therefore, we can reduce power consumption for writing data using the electric-field-assisted magnetization reversal.[14]

The ferromagnetic semiconductor GaMnAs is one of the most suitable material systems for the proof-of-concept study of this device, because the current-in-plane spin-valve effect has been observed in GaMnAs / GaAs / GaMnAs heterostructures,[15–18] and the channel resistance of single GaMnAs layers can be modulated by the gate electric field if the thickness of the GaMnAs layer is sufficiently thin (~ 5 nm).[19] The heterostructure used in this study is composed of GaAs (1.5 nm) / $Ga_{0.94}Mn_{0.06}As$ (5 nm) / GaAs (9 nm) / $Ga_{0.94}Mn_{0.06}As$ (5 nm) / $In_{0.17}Al_{0.83}As$ (500 nm) / GaAs (100 nm) grown on semi-insulating (001)-oriented GaAs substrates by low-temperature molecular beam epitaxy. The schematic illustration of the heterostructure studied here is shown in Fig. 1(b). The growth temperatures of these layers are, from the top to the bottom, 185, 185, 170, 190, 430, and 500 °C, respectively. The reflection high energy electron diffraction (RHEED) showed a clear 1×2 surface reconstruction during the growth of $Ga_{0.94}Mn_{0.06}As$, indicating 2-dimensional growth and no obvious segregation of hexagonal MnAs. The lattice-relaxed $In_{0.17}Al_{0.83}As$ buffer layer introduces tensile strain in the $Ga_{0.94}Mn_{0.06}As$ / GaAs / $Ga_{0.94}Mn_{0.06}As$ layers, and thus the magnetic easy axes of the two $Ga_{0.94}Mn_{0.06}As$ layers are perpendicular to the film plane.[20] The sample was fabricated into Hall-bar structures with the width of 50 μm and the length of 200 μm by photolithography and chemical wet etching to measure the spin-dependent transport properties (Fig. 1(c)). Then, a 50-nm-thick $HfO_2$ layer was deposited as a gate insulator by atomic layer deposition at the substrate temperature of 100 °C and contact holes were opened. Finally, after the electron-beam deposition of a 50-nm-thick Au layer, electrodes were formed by photolithography and chemical wet etching of the Au layer. We connected the gold wires to the electrodes using Indium solder. We measured the spin-dependent transport properties of this device under various gate voltages $V_G$ and external magnetic fields $H$ applied perpendicular to the film plane. The longitudinal and transverse voltages ($V_{xx}$ and $V_{xy}$, respectively) were measured simultaneously at a constant direct current $I$ of 10 µA. Because the ordinary Hall effect is negligibly small due to the large carrier concentration in GaMnAs, the Hall resistance ($R_{Hall} = V_{xy}/I$) is dominated by the anomalous Hall effect. This enables us to detect the magnetization configuration through measuring $R_{Hall}$.

Figure 2(a) shows $R_{Hall}$ as a function of $\mu_0 H$ at various temperatures $T$ ranging from 3.8 K to 60 K at $V_G = 0$ V, where $\mu_0$ is the permittivity of a vacuum. Clear hysteresis loops and double step features are seen at $T$ up to 46 K. These results indicate that both $Ga_{0.94}Mn_{0.06}As$ layers are ferromagnetic below 46 K. The steps observed in the $R_{Hall} - \mu_0 H$ curves in Fig. 2(a) correspond to the antiparallel magnetization configuration. The sharp double step feature in the $R_{Hall} - \mu_0 H$ curve disappeared when $T > 50$ K because the Curie temperatures of both $Ga_{0.94}Mn_{0.06}As$ layers are around 50 K. Figure 2(b) shows $R_{Hall}$ and $R_{sheet}$ (= $V_{xx}/I$) as a function of $\mu_0 H$ at 36 K and $V_G = 0$ V. The abrupt spike-like peaks in $R_{sheet}$ observed at the coercive fields of each layer (-7.95 mT, -7.3 mT, 7.25 mT, and 7.8 mT, denoted by black inverted triangles) do not originate from the spin-valve effect, but probably from the presence of a domain wall. It is known that abrupt jumps occur in $R_{sheet}$ due to the presence of a domain wall when the magnetization is reversed in single GaMnAs layers with a tensile strain.[21–24] However, these spike-like peaks are not essential in our device. As clearly seen in the magnified plots of $R_{Hall} - \mu_0 H$ and $R_{sheet} - \mu_0 H$ in Fig. 2(c), $R_{sheet}$ is larger in the range of $\mu_0 H$ from -7.7 mT to -7.3 mT, where the magnetization configuration is antiparallel, than that in the parallel magnetization configuration. This is the evidence of the spin-valve effect in the GaMnAs / GaAs / GaMnAs trilayer structure.[25–27] The MR ratio, which is defined as $(R_{sheet}(H) - R_{sheet}(H = 0)) / R_{sheet}(H = 0) \times 100\%$, at 36 K reached 0.17% at $V_G = 0$ V, where $R_{sheet}(H)$ represents $R_{sheet}$ at an external magnetic field of $H$.

The magnetic properties can be modulated by $V_G$ in our CIP-SV-FET. Figures 3(a) – 3(f) show $R_{Hall}$ (black solid line, left axis) and $R_{sheet}$ (blue circles, right axis) as a function of $\mu_0 H$ from -8.5 mT to -6.5 mT under the various gate voltages from 0 V to 25 V at 36 K. The coercive force of the top $Ga_{0.94}Mn_{0.06}As$ layer ($H_{top}$) (denoted by the green inverted triangles in Figs. 3(a) – 3(f)) was changed from -7.8 mT to -7.1 mT when $V_G$ was changed from 0 V to 25 V. Because only the top $Ga_{0.94}Mn_{0.06}As$ layer is subject to the gate electric field, the coercive force of the bottom $Ga_{0.94}Mn_{0.06}As$ layer ($H_{bottom}$) remained almost unchanged at -7.3 mT as denoted by the red inverted triangles in Figs. 3(a) – 3(f). The change in the coercive force of the top GaMnAs layer by the gate electric field is explained by the change of the nucleation field.[11] The shape of the $R_{sheet}$ – $\mu_0 H$ curve is changed in a little bit complicated manner by $V_G$. At $V_G = 0$ V, $|H_{top}| > |H_{bottom}|$, and normal positive spin-valve effect was observed. With increasing $V_G$ from 0 V to 10 V, $|H_{top}|$ decreases and becomes close to $|H_{bottom}|$, and $R_{sheet}$ still shows normal positive MR. However, when $V_G > 20$ V and $|H_{top}|$ becomes smaller than $|H_{bottom}|$, $R_{sheet}$ in the antiparallel magnetization configuration becomes smaller than that in the parallel magnetization configuration. This is because $R_{sheet}$ in the anti-parallel configuration is expressed by the sum of the normal positive MR and the negative peak induced by the domain wall. As can be seen in Figs. 3(a) – 3(f), the negative peak induced by the domain wall moves toward the smaller $|\mu_0 H|$ region with increasing $V_G$ (-8.1 mT − -7.8 mT at $V_G$ = 0 V, -7.3 mT − -7.0 mT at $V_G$ = 25 V). This peak overlaps with the anti-parallel magnetization region (i.e. between the red and green arrows) and thus hides the normal positive MR when $V_G$ = 20 V and 25 V. The electrical properties can also be controlled by $V_G$ as shown in Fig. 3(g). $R_{sheet}$ was changed by 14% (42.3 kΩ → 48.4 kΩ) by $V_G$ at 36 K. These results indicate that the electrical and magnetic properties can be changed by $V_G$ in our CIP-SV-FET.

Because the magnetic properties can be modulated by $V_G$ in our CIP-SV-FET, we can use electric-field-assisted magnetization reversal in this device. To demonstrate it, we measured the response of $R_{Hall}$ to the gate voltages at a fixed $H$ (Fig. 4(a)). Before the measurements, the magnetizations were aligned parallel by applying $\mu_0 H$ of +1 T. Then, $\mu_0 H$ was swept back to -7.6 mT under $V_G$ of -25 V. Because the coercive force of the top GaMnAs layer is -7.4 mT when $V_G$ = -25 V (see the inset of Fig. 4(a)), the magnetization of the top GaMnAs layer is reversed by this process and the magnetizations become antiparallel. When time $t$ < 20 s, we kept $V_G$ = -25 V and $R_{Hall}$ = -0.15 kΩ, which indicates that the magnetization configuration remains antiparallel. At $t$ ~ 20 s, $V_G$ was changed from -25 V to 25 V. Following this, the hysteresis loop of the top $Ga_{0.94}Mn_{0.06}As$ layer expressed by the blue curve is changed to the red curve in the inset of Fig. 4(a). Thus, the magnetization of the top $Ga_{0.94}Mn_{0.06}As$ layer is reversed, and the magnetizations become parallel and $R_{Hall}$ decreased to -0.8 kΩ. At $t$ ~ 40 s, we switched $V_G$ from 25 V to -25 V. At this time, the magnetic static energy at the top GaMnAs layer is already lowest because the magnetization direction is parallel at the magnetic field. Thus, the hysteresis loop is not switched back to the blue curve in the inset of Fig. 4(a), and the magnetization configuration remains parallel. We can also reverse the magnetization configuration from the antiparallel state to the parallel state by following the same procedure if $\mu_0H$ is swept back to -7.3 mT under $V_G$ of -25 V. In this way, by utilizing the electric field, we can reduce the magnetic field required for magnetization reversal (-8 mT → -7.6 mT), which contributes to the power reduction for writing data. These results mentioned above indicate that electric-field-assisted magnetization reversal can be used in our spin device.

We show that the characteristics of the current $I$ *vs*. the voltage $V$ between the source and the drain electrodes can be controlled both by the magnetization configuration and the gate voltage. In Fig. 4(b), the red (blue) solid and dashed lines represent $I$ as a function of $V$ at $V_G$ = 25 V (-25 V) in the parallel and antiparallel magnetization configurations, respectively. The current under the antiparallel magnetization configuration was measured at $\mu_0H$ = 0 mT in the minor loop shown in the inset of Fig. 4(b). The current modulation by the magnetization configuration and the gate voltage is the basic operation of spin MOSFETs. Thus, our results suggest that the CIP-SV-FET can be used as an alternative type of the spin MOSFETs.

In summary, we have investigated the spin-dependent transport properties of the CIP-SV-FET. The resistance was modulated both by magnetization configuration (0.17%) and gate electric field (14%) at 36 K. The electric-field-assisted magnetization reversal was also demonstrated in this device. These results indicate that the CIP-SV-FET is a hopeful candidate for an alternative type of a spin MOSFET.

This work was supported by Grants-in-Aid for Scientific Research, the Project for Developing Innovation Systems of MEXT, and Spintronics Research Network of Japan. Part of this work was carried out under the Cooperative Research Project Program of RIEC, Tohoku University. T. Kanaki acknowledges the financial support

FIG. 1. (Color online) (a) Schematic illustration of a CIP-SV-FET. (b)(c) Schematic illustration of the GaMnAs-based heterostructure (b) and the Hall-bar CIP-SV-FET device used in this study (c). The width (length) of the Hall bar is 50 μm (200 μm). With a constant direct current $I$ (10 μA), the longitudinal (transverse) voltage, defined as $V_{xx}$ ($V_{xy}$), is measured under a magnetic field $\mu_0 H$ applied perpendicular to the film plane and various gate voltages $V_G$.

FIG. 2. (Color online) (a) Hall resistance $R_{Hall}$ as a function of the external magnetic field $\mu_0 H$ applied perpendicular to the film plane at various temperatures. (b)(c) Hall resistance $R_{Hall}$ and sheet resistance $R_{sheet}$ as a function of $\mu_0 H$ applied perpendicular to the film plane at 36 K (b) and those in the range of $\mu_0 H$ from -9 mT to -6.5 mT (c). In (a) – (c), the gate voltage $V_G$ is 0 V. The black triangles represent the coercive fields of each layer in (b) and the magnetization configurations (white arrows) are shown in (b) and (c). The parallel (antiparallel) regions are colored by pink (blue).

FIG. 3. (Color online) (a) – (f) Hall resistance $R_{Hall}$ (left axis) and sheet resistance $R_{sheet}$ (right axis) as a function of the external magnetic field $\mu_0 H$ under various gate voltages $V_G$ from 0 V to 25 V. The data were obtained at 36 K. The green (red) triangles represent $H_{top}$ ($H_{bottom}$). (g) Sheet resistance $R_{sheet}$ as a function of $V_G$ with a bias current of 10 μA at 36 K and 0 mT when the magnetization configuration is parallel.

FIG. 4. (Color online) (a) Response of the Hall resistance $R_{Hall}$ to the gate voltage at 36 K. Here, $t$ is time. The red (blue) circles represent $R_{Hall}$ at $V_G$ = 25 V (-25 V). The external magnetic field $\mu_0 H$ was fixed at -7.6 mT. The inset shows $R_{Hall}$ as a function of $\mu_0 H$ at $V_G = \pm 25$ V. (b) Current $I$ as a function of the bias voltages $V$ at 3.8 K and 0 mT. The red (blue) solid and dashed lines are the data obtained when $V_G = 25$ V (-25 V) in the parallel (solid curves) and antiparallel (broken curves) magnetization configurations. The black solid (dotted) curve in the inset is the major (minor) loop of $R_{Hall}$ at 3.8 K.

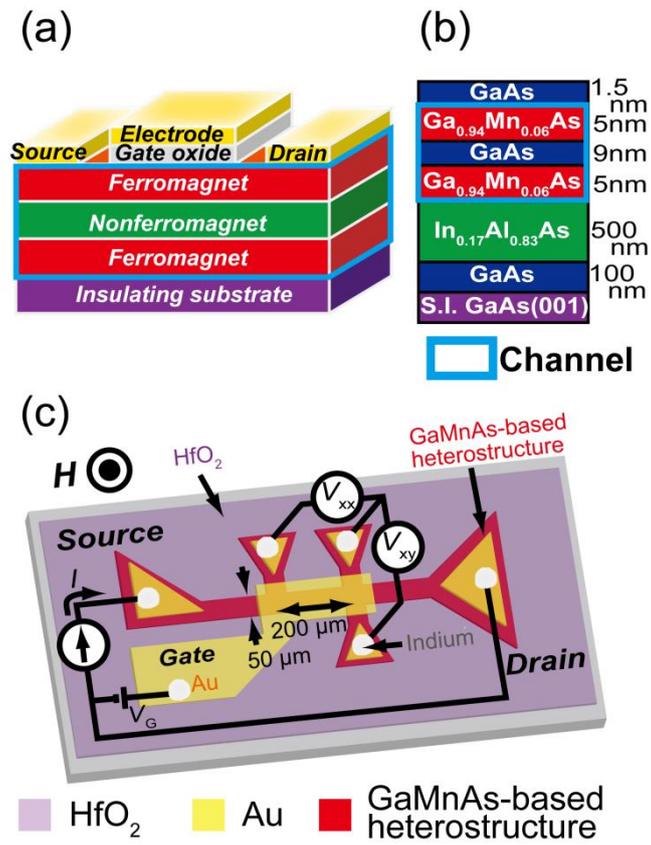

Fig. 1. Kanaki *et al*.

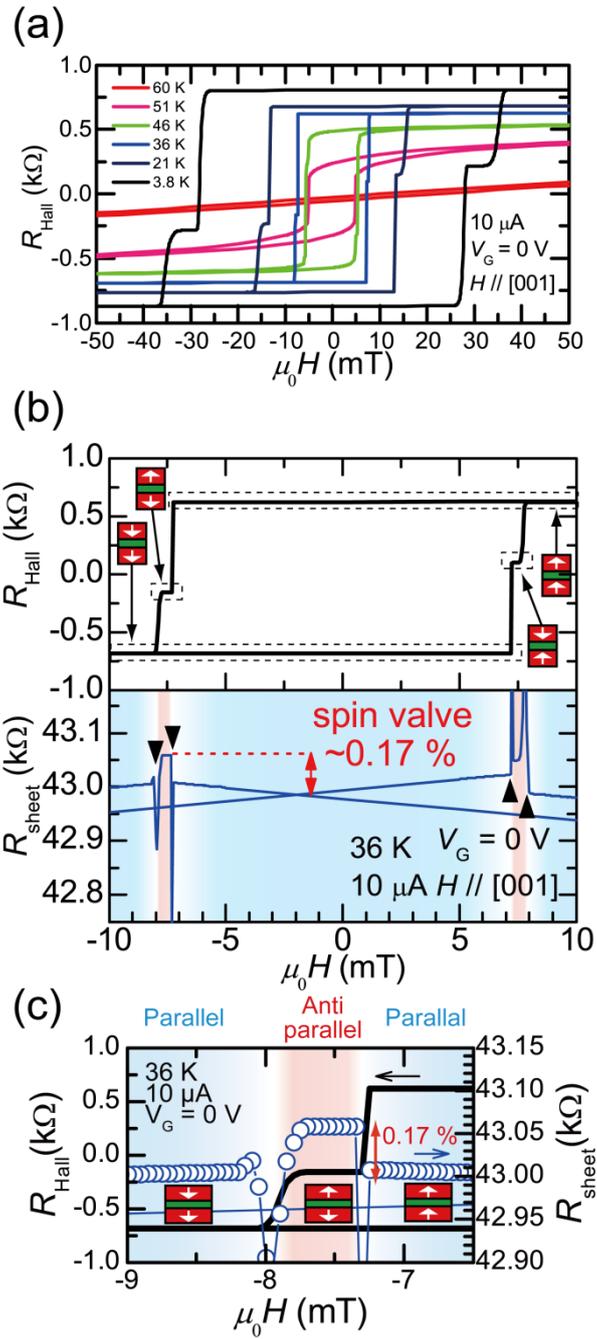

Fig. 2. Kanaki *et al*.

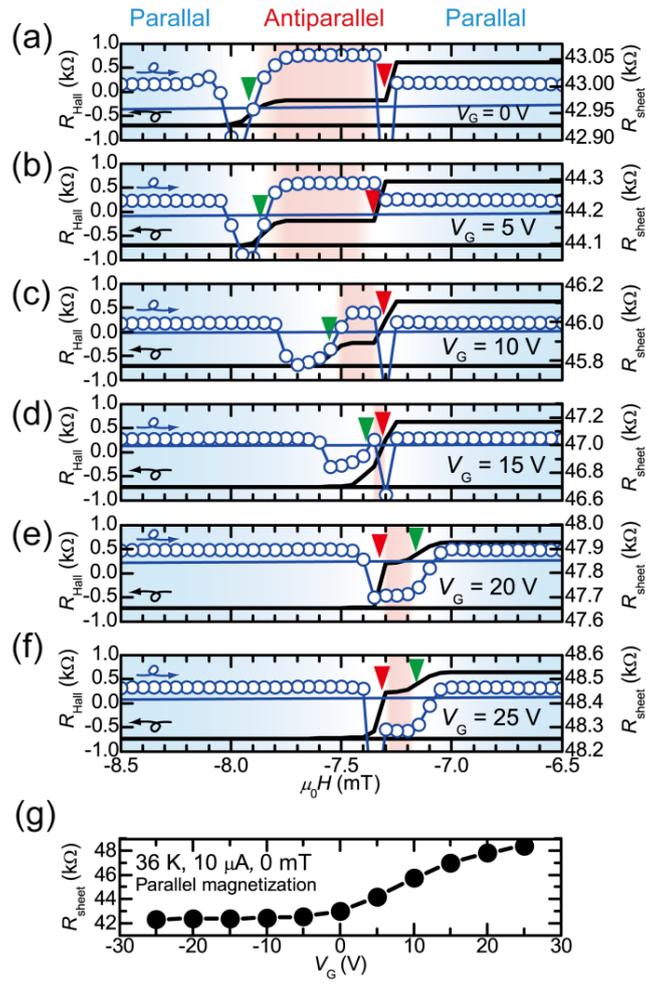

Fig. 3. Kanaki *et al*.

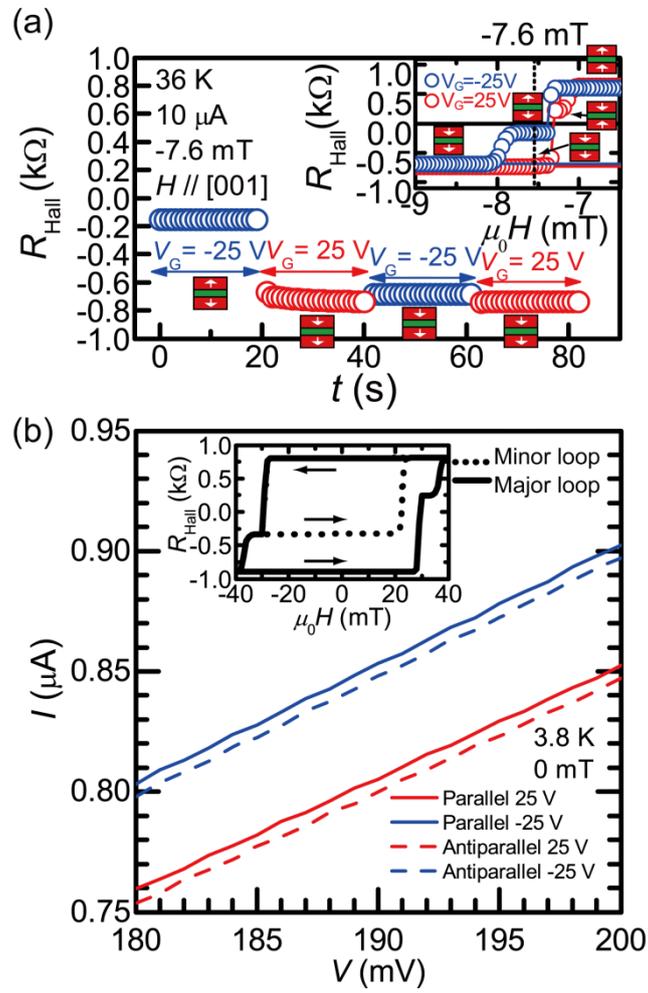

Fig. 4. Kanaki *et al*.